\documentclass[preprint]{aastex}

\def\Msol{\thinspace\hbox{$\hbox{M}_{\odot}$}}

\def\a4{\hsize 17.0cm \vsize 25.cm}

\shorttitle{Evolution of SSC Winds with Strong Cooling}
\shortauthors{ W\"unsch et al.}

\begin{document}

\title{Evolution of Super Star Cluster Winds with Strong Cooling}

\author{Richard W\"unsch}
\affil{Astronomical Institute, Academy of Sciences of the Czech
Republic, Bo\v{c}n\'\i\ II 1401, 141 31 Prague, Czech Republic}
\author{Sergiy Silich}
\affil{Instituto Nacional de Astrof\'\i sica Optica y
Electr\'onica, AP 51, 72000 Puebla, M\'exico}
\author{Jan Palou\v{s}}
\affil{Astronomical Institute, Academy of Sciences of the Czech
Republic, Bo\v{c}n\'\i\ II 1401, 141 31 Prague, Czech Republic}
\author{Guillermo Tenorio-Tagle}
\affil{Instituto Nacional de Astrof\'\i sica Optica y
Electr\'onica, AP 51, 72000 Puebla, M\'exico}
\and
\author{Casiana Mu\~noz-Tu\~n\'on}
\affil{Instituto de Astrof\'\i sica de Canarias, c/ V\'\i a L\'actea s/n, E-38205, La Laguna,
Tenerife and Departamento de Astrofisica, Universidad de La Laguna, E-38205, La
Laguna, Tenerife, Spain}

\begin{abstract}

We study the evolution of Super Star Cluster (SSC) winds driven by stellar winds
and supernova (SN) explosions. Time-dependent rates at which mass and energy are
deposited into the cluster volume, as well as the time-dependent chemical
composition of the re-inserted gas, are obtained from the population synthesis
code Starburst99. These results are used as input for a semi-analytic code which
determines the hydrodynamic properties of the cluster wind as a function of
cluster age. Two types of winds are detected in the calculations. For the
quasi-adiabatic solution, all of the inserted gas leaves the cluster in the form
of a stationary wind. For the bimodal solution, some of the inserted gas becomes
thermally unstable and forms dense warm clumps which accumulate inside the
cluster. We calculate the evolution of the wind velocity and energy flux and
integrate the amount of accumulated mass for clusters of different mass, radius
and initial metallicity. We consider also conditions with low heating
efficiency of the re-inserted gas or mass loading of the hot thermalized plasma
with the gas left over from star formation. We find that the bimodal regime and
the related mass accumulation occur if at least one of the two conditions above
is fulfilled. 

\end{abstract}


\section{Introduction}
\label{sec:intro}

Super star clusters are young compact objects observed in many starburst
and interacting galaxies in a variety of wavelengths 
\citep{
1992AJ....103..691H,
1993AJ....106.1354W,
1995ApJ...446L...1O,
2005ApJ...619..270M,
2006MNRAS.370..513S,
2007ApJ...668..168G,
2008A&A...492....3G,
2011ApJ...729..111W}. 
With masses $10^5 - 10^7$~M$_\odot$ and ages $\lesssim 10^7$~yr they are expected to
include large numbers of massive stars which lose substantial fractions of their
mass via stellar winds and supernova explosions.

\citet[hereafter CC85]{1985Natur.317...44C} studied the hydrodynamics of the gas
re-inserted by massive stars into the cluster interior using an adiabatic
spherically-symmetric model. They assumed that the mechanical energy of stellar
winds and supernovae ejecta is thermalized in random collisions and the gas
within the cluster is heated up to $\sim 10^7$~K. The resulting high pressure
drives the cluster wind for which CC85 found a stationary hydrodynamic solution.
They assumed that the mass and the thermal energy are inserted uniformly at
rates $\dot{M}_\mathrm{SC}$ and $L_\mathrm{SC}$, respectively, into a sphere
(cluster) of radius $R_\mathrm{SC}$. They showed, that under such
assumptions, a stationary wind can only be obtained if the flow velocity equals
zero at the cluster center and reaches the sound speed exactly at the cluster
border. Super star cluster winds were studied further using analytical and
numerical models by many authors including \citet{ 2000ApJ...536..896C,
2001ApJ...559L..33R,2003ApJ...590..791S,2006ApJ...643..186T}.

It was found by \citet{2004ApJ...610..226S} that the adiabatic
approximation becomes inadequate for very massive and compact clusters. The
authors showed that the stationary solution of the cluster wind does not
exist for clusters with $L_\mathrm{SC}$ larger than a critical value
$L_\mathrm{crit}$. This is because the total energy input rate, $L_\mathrm{SC}$,
is proportional to the cluster stellar mass, $M_\star$, while the energy losses
from the hot gas due to radiation are proportional to $M_\star^2$ (since cooling
is proportional to the second power of the gas density which is proportional to
$M_\star$). \citet{2004ApJ...610..226S} showed how $L_\mathrm{crit}$ depends
on the star cluster parameters and \citet{2007A&A...471..579W} founded an
approximate analytical formula for $L_\mathrm{crit}$.

Clusters with $L_\mathrm{SC} > L_\mathrm{crit}$ were studied by means of 1D
hydrodynamic simulations by \citet{2007ApJ...658.1196T}, who showed that
such clusters evolve in the bimodal hydrodynamic regime. In such a case, the
cluster is divided by the stagnation radius, $R_\mathrm{st}$, into two
qualitatively different regions. The stationary wind solution still exists in
the outer region $r > R_\mathrm{st}$, with the wind velocity being zero at
$R_\mathrm{st}$ and reaching the sound speed at $R_\mathrm{SC}$. In the region
$r < R_\mathrm{st}$, on the other hand, the thermal instability sets in and
random parcels of gas cool down to $\sim 10^4$~K (further cooling is prevented
by the intense stellar radiation). Consequently, the warm regions are compressed
into dense clumps by repressurizing shocks driven by the surrounding hot gas.
Clusters in the bimodal regime were studied further by
\citet{2008ApJ...683..683W} who used 2D hydrodynamics to follow the clump
formation, and to estimate the fraction of the re-inserted matter which leaves
the cluster as a wind and the fraction which accumulates inside the stagnation
radius and possibly leads to secondary star formation
\citep{2005ApJ...628L..13T}.

It was suggested that two-component supersonic recombination line profiles often
detected in young and massive SSCs \citep{2007ApJ...668..168G,
2008A&A...489..567B, 2007AJ....133..757H} and compact dense HII regions
overlapping young SSCs \citep{2006MNRAS.370..513S} may present the observational
manifestation for such bimodal regime \citep{2010ApJ...708.1621T,
2007ApJ...669..952S, 2009ApJ...700..931S}. In both cases the calculations
require the shocked gas temperature to be lower than that predicted by the CC85 model
as it is also the case when the model predicted diffuse X-ray emission is
compared to the observed values \citep{2003MNRAS.339..280S}. Two different
processes which may decrease the intercluster gas temperature have been
discussed in the literature: the efficiency with which the kinetic energy of
stellar winds and SNe is thermalized, and the additional mass loading into the
hot gas inside the cluster \citep{2003MNRAS.339..280S, 2004A&A...424..817M, 2007A&A...471..579W,
2007ApJ...669..952S, 2009ApJ...700..931S, 2010ApJ...711...25S}. In this work we
do not discuss details related to those two processes, however, we introduce two
free parameters $\eta_\mathrm{he}$ and $\eta_\mathrm{ml}$ and show how the
results depend on their values.

Previous works on clusters in the bimodal regime use the energy and mass
deposition rates $L_\mathrm{SC}$ and $\dot{M}_\mathrm{SC}$ as free parameters.
In this work, we calculate time-dependent $L_\mathrm{SC}(t)$ and
$\dot{M}_\mathrm{SC}(t)$ using the stellar population synthesis code Starburst99
\citep{1999ApJS..123....3L} for a cluster with a given stellar mass, $M_\star$,
and initial stellar metallicity, $Z_0$. Subsequently, we insert
$L_\mathrm{SC}(t)$ and $\dot{M}_\mathrm{SC}(t)$ into our semi-analytic code to
determine the evolutionary properties of the cluster wind. We also calculate
whether the cluster spends some time in the bimodal regime and estimate the
amount of re-inserted gas which becomes thermally unstable and accumulates
inside the cluster. The Starburst99 code also provides us with the time
evolution of the re-inserted gas chemical composition. The chemical
composition is an important parameter as the cooling rate depends on it. This
work effectively replaces the three functions of time $L_\mathrm{SC}(t)$,
$\dot{M}_\mathrm{SC}(t)$ and $Z(t)$ (metallicity of the cluster wind), with the
two constant parameters: mass of the star cluster $M_\star$ and its initial
metallicity $Z_0$.

The paper is organized as follows: in \S\ref{sec:model} we describe the
semi-analytic code used for the calculation of the cluster wind and the way
how it utilizes results of the Starburst99 code. In \S\ref{sec:results} we show
results for a reference model with $M_\star = 10^6$~M$_\odot$ and $R_\mathrm{SC}
= 3$~pc (\S\ref{ssec:proto}) and give the dependence of results on the cluster
mass, the cluster radius and the initial stellar metallicity (\S\ref{ssec:dep}).
In \S\ref{sec:conclusions} we summarize our conclusions.

\begin{figure}
\plotone{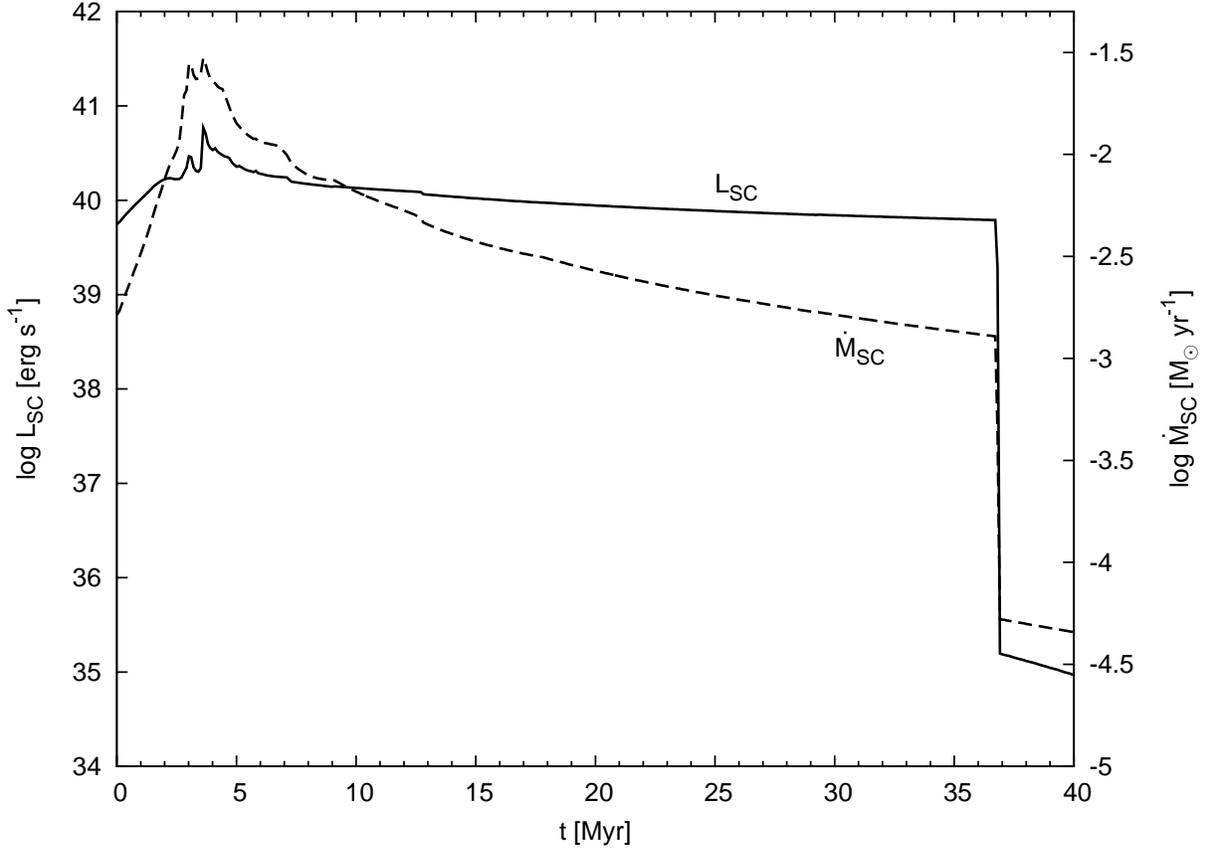}
\caption{Energy (solid line, left y-axis) and mass (dashed line, right y-axis)
deposition rates calculated by the Starburst99 code for the reference model
$M_\star = 10^6$~M$_\odot$, $R_\mathrm{SC} = 3$~pc, $Z_0 = $Z$_\odot$,
$\eta_\mathrm{he} = 1$ and $\eta_\mathrm{ml} = 0$.
}
\label{fig:depos}
\end{figure}

\begin{figure}
\plotone{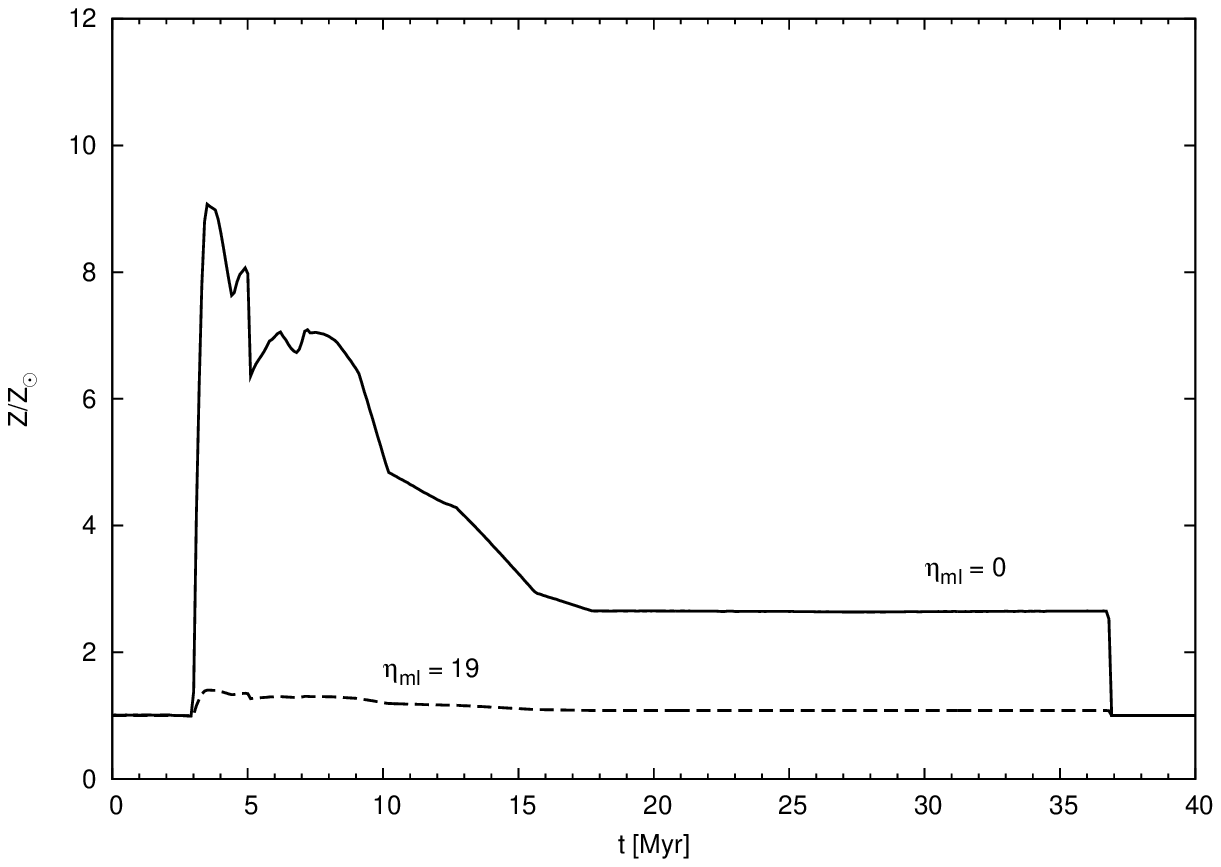}
\caption{The evolution of the metallicity of the hot shocked gas inside the
cluster with $Z_0 = Z_\odot$.
The solid and dashed lines show the metallicity without ($\eta_\mathrm{ml} = 0$)
and with ($\eta_\mathrm{ml} = 19$) mass loading, respectively.}
\label{fig:chem}
\end{figure}

\section{The cluster wind}
\label{sec:model}

In this section we specify the assumptions used in the semi-analytic model of
the cluster wind and formulate its basic equations. We also describe how the
model equations are integrated and the properties of the bimodal solution (e.g.
$R_\mathrm{st}$, $L_\mathrm{crit}$ and $\dot{M}_\mathrm{wind}$) determined. 
Finally, we describe how the wind model utilizes the output from the Starburst99
code.

\subsection{Assumptions and basic equations}

We consider a spherical cluster of radius $R_\mathrm{SC}$ whose stars deposit
mass and energy at rates $\dot{M}_\mathrm{SC}$ and $L_\mathrm{SC}$,
respectively. We assume, similar to CC85, that mutual collisions of stellar
winds and SNe ejecta, and collisions with gas left over from the formation of
the first stellar generation and with gas returned by pre-main sequence stars
via outflows, result in the production of hot gas which occupies most of the
cluster volume. Therefore, we model these processes by inserting mass and energy
uniformly into the whole cluster volume with deposition rate densities $q_m$ and
$q_e$. In order to account for the uncertainties related to the thermalization
of the mechanical energy of the inserted gas we introduce the heating
efficiency, $\eta_\mathrm{he} \in (0,1)$, denoting the fraction of the
mechanical energy of stellar winds and SN ejecta which is converted into the
thermal energy of the hot gas. Furthermore, we assume that a substantial amount
of gas was left over after the formation of the first generation of stars and
that this gas may evaporate and be dispersed into the hot gas. Another
contribution to the mass of hot gas inside the cluster comes from outflows of
pre-main sequence stars which are not included in the Starburst99 code. Indeed,
the mass left over from star formation and the T~Tauri multiple outflows, such
as the jet from RW Aurigae \citep{1996A&A...310..309B}, should make a
substantial contribution to the mass available for mass loading. We describe
these processes by the mass loading factor, $\eta_\mathrm{ml} \in (0,\infty)$,
which gives the amount of the loaded mass relative to $\dot{M}_\mathrm{SC}$. The
total mass injection rate then is $\dot{M}_\mathrm{in} = \dot{M}_\mathrm{SC} +
\eta_\mathrm{ml} \dot{M}_\mathrm{SC} = (1+\eta_\mathrm{ml})\dot{M}_\mathrm{SC}$.
It is assumed that the metallicity of the loaded gas is the same as the initial
stellar metallicity, $Z_0$. 

The spherically symmetric hydrodynamic equations describing the wind flow are
\citep[CC85,][]{2004ApJ...610..226S}
\begin{equation}
\label{basic_con}
\frac{1}{r^2}\frac{d}{dr}(\rho u r^2) = q_m
\end{equation}

\begin{equation}
\label{basic_mom}
\rho u \frac{du}{dr} = - \frac{dP}{dr} - q_m u
\end{equation}

\begin{equation}
\label{basic_ener}
\frac{1}{r^2} \frac{d}{dr}\left[
\rho u r^2 \left( \frac{u^2}{2} + \frac{\gamma}{\gamma-1}\frac{P}{\rho} \right)
\right] = q_e - Q
\end{equation}
where $\gamma$ is the adiabatic index and $\rho$, $u$ and $P$ are wind density,
velocity and pressure, respectively. Mass and energy deposition rate densities
$q_m$ and $q_e$ are
\begin{equation}
\label{eq:qmqe}
\begin{array}{rcl}
q_m & = & \frac{3(1+\eta_\mathrm{ml})\dot{M}_\mathrm{SC}}{4\pi
R_\mathrm{SC}^3}\\
q_e & = & \frac{3\eta_\mathrm{he}L_\mathrm{SC}}{4\pi R_\mathrm{SC}^3}
\end{array}
\end{equation}
for $r < R_\mathrm{SC}$ and $q_m=q_e=0$ for $r > R_\mathrm{SC}$. The energy
equation (\ref{basic_ener}) includes the cooling term $Q = n_i n_e \Lambda(T,
Z)$ where $n_i = n_e = \rho/\mu_i$ are the ion and electron number densities,
$Z$ is the gas metallicity and $\Lambda(T,Z)$ is a cooling
function tabulated by \citet{1995MNRAS.275..143P}. We
use $\mu_i = 14/11 m_\mathrm{H}$ neglecting the
contribution of heavy elements.

Several interesting properties may be derived directly from equations
(\ref{basic_con}) -- (\ref{basic_ener}) \citep[see][for
details]{2004ApJ...610..226S}. Firstly, the stationary solution exists only if
the wind velocity, $u$, reaches the sound speed exactly at the cluster border.
Secondly, a relation between the temperature $T_\mathrm{st}$ and the density
$\rho_\mathrm{st}$ at the stagnation radius can be derived 
\begin{equation} 
\label{rhost} 
\rho_\mathrm{st} = \left[\frac{q_e - q_m c^2_\mathrm{st} / (\gamma - 1)}
{\Lambda(T_\mathrm{st},Z)}\right]^{1/2} ,
\end{equation}
where $c_\mathrm{st}$ is the sound speed at the stagnation radius. 
Furthermore, it has been shown by \citet{2007ApJ...658.1196T} that if the
cluster is in the bimodal regime, i.e. if $R_\mathrm{st} > 0$, the pressure at the
stagnation radius $P_\mathrm{st} = (k\rho_\mathrm{st}T_\mathrm{st})/\mu_a$
reaches the maximum value $P_\mathrm{max} = \max
(P_\mathrm{st}(T_\mathrm{st}))$, where $k$ denotes the Boltzmann constant and
$\mu_a = 14/23 m_\mathrm{H}$ is the mean mass per particle.

\subsection{Integration procedure}

The wind solution is found by the following procedure. At first, it is assumed
that $R_\mathrm{st} = 0$ and an attempt to find $T_\mathrm{st}$ is made.
Equations (\ref{basic_con}) -- (\ref{basic_ener}) are repeatedly numerically
integrated from $r = 0$ to $R_\mathrm{SC}$ with $T_\mathrm{st}$ varying in the
interval $(0, T_a)$ where $T_a = (\gamma-1)\mu_a q_e/(\gamma kq_m)$ is the
adiabatic wind central temperature. The central density $\rho_\mathrm{st}$ is
calculated from equation~(\ref{rhost}). Then, the bisection method is used to
find $T_\mathrm{st}$ for which the sonic radius $R_\mathrm{son}$ (defined as
$u(R_\mathrm{son}) = c_s(R_\mathrm{son})$) is equal to $R_\mathrm{SC}$.

If this attempt fails (i.e. no initial conditions at $r = 0$ for which
$R_\mathrm{son} = R_\mathrm{SC}$ exist), it implies that $R_\mathrm{st} > 0$ and
the cluster is in the bimodal regime. In such a case, 
the value of $T_\mathrm{st}$ is defined by the condition that the function
$P_\mathrm{st}(T_\mathrm{st})$ has its maximum $P_\mathrm{max}$ \citep{2007ApJ...658.1196T}. Therefore,
the temperature at the stagnation radius is found using the golden section method and it is used as
the initial condition for integrating equations~(\ref{basic_con}) --
(\ref{basic_ener}). Then, similarly as in the previous case, $R_\mathrm{st}$ is
varied and the bisection method is used to find the solution which satisfies the
condition $R_\mathrm{son} = R_{SC}$.

Once all the initial conditions ($R_\mathrm{st}$, $\rho_\mathrm{st}$ and
$T_\mathrm{st}$) are known, radial profiles of the wind density $\rho$, velocity
$u$ and temperature $T$ can be obtained by integrating
equations~(\ref{basic_con}) -- (\ref{basic_ener}) in the interval
$(R_\mathrm{st}, 10~R_\mathrm{SC}$. The semi-analytic model is unable to describe the
inner thermally unstable region with $r < R_\mathrm{st}$. However, 2D hydrodynamic
simulations \citep{2008ApJ...683..683W} have shown that the temperature and the
density of the hot gas in this region are close to uniform and stay constant with time. The deposition of
mass into this region is balanced by the formation of dense warm clumps 
which tend to accumulate in this region. Therefore, we assume that
the hot gas in the central region $r < R_\mathrm{st}$ has zero velocity, uniform density
$\rho_\mathrm{st}$ and temperature $T_\mathrm{st}$, and that all gas
inserted into this region accumulates there. Finally, the critical luminosity
$L_\mathrm{crit}$ is determined by repeating the above procedure and searching
for the lowest mechanical luminosity $L_\mathrm{SC}$ for which $R_\mathrm{st} >
0$.

\subsection{Starburst99 outputs used in the wind model}

The stellar population synthesis code Starburst99 \citep{1999ApJS..123....3L}
calculates a set of stellar evolution models for a given population of stars and
determines their collective properties. In this work, the total mass loss rate
from stellar winds and SNe type II ejecta is used as the mass deposition rate,
$\dot{M}_\mathrm{SC}$, and the total stellar wind and SNe ejecta power as the
energy deposition rate, $L_\mathrm{SC}$. All Starburst99 simulations used in
this work are set up with the following parameters: star formation is
instantaneous with the fixed stellar mass $M_\star$; the standard Kroupa Initial
Mass Function \citep{2001MNRAS.322..231K} with two power-laws ($dN/dm \sim
m^{-1.3}$ between $0.1$ and $0.5$~\Msol and $dN/dm \sim m^{-2.3}$ between $0.5$
and $100$~\Msol) is used; the supernova cut-off mass is equal to $8$~\Msol;
stellar evolutionary tracks are Geneva with high mass loss; and the wind model
is evolutionary (see \citealp{1992ApJ...401..596L} for details). The evolution
of $\dot{M}_\mathrm{SC}$ and $L_\mathrm{SC}$ for the reference model (see
\S\ref{ssec:proto}) is shown in Figure~\ref{fig:depos}. We have followed the
first $40$~Myr of the cluster evolution. This period is long enough to cover the
life time of all massive stars even in cases with initial stellar metallicities,
$Z_0$, different than $Z_\odot$, discussed in \S\ref{ssec:dep}. We do not
consider here the period after the last massive star explodes (this moment is
visible as a sudden drop of $\dot{M}_\mathrm{SC}$ and $L_\mathrm{SC}$ at
$37$~Myr in Figure~\ref{fig:depos}).

Starburst99 also provides the chemical composition of the re-inserted matter by
specifying mass loss rates for: H, He, C, N, O, Mg, Si, S and Fe. Thus one can
calculate the injection rate for seven elements heavier than H and He
\begin{equation}
\dot{M}_\mathrm{metals} = \sum_{j=C}^{Fe} \dot{M}_j
\end{equation}
where $\dot{M}_j$ is the mass deposition rate of the $j$-th element. It is assumed that the
injected gas is rapidly mixed with the mass loaded gas. The
metallicity of the cluster wind, $Z$, used in equation (\ref{basic_ener}) is
\begin{equation}
\label{eq:zmix}
Z =
\frac{\dot{M}_\mathrm{metals}+\eta_\mathrm{ml}Z_0\dot{M}_\mathrm{SC}}
{(1+\eta_\mathrm{ml})\dot{M}_\mathrm{SC}}
\ .
\end{equation}
The evolution of $Z$ in the cluster with $Z_0=Z_\odot$ for different values of
$\eta_\mathrm{ml}$ is shown in Figure~\ref{fig:chem}. Taken together, the model
utilizing Starburst99 results includes five parameters: $M_\star$, $R_\mathrm{SC}$,
$Z_0$, $\eta_\mathrm{he}$ and $\eta_\mathrm{ml}$. The semi-analytic wind model
on its own includes six parameters: $\dot{M}_\mathrm{SC}$, $L_\mathrm{SC}$, $Z$,
$R_\mathrm{SC}$, $\eta_\mathrm{he}$ and $\eta_\mathrm{ml}$. Here we assume that
the first three of them ($\dot{M}_\mathrm{SC}$, $L_\mathrm{SC}$, $Z$) are
functions of the star cluster age. We keep the heating efficiency,
$\eta_\mathrm{he}$, and the mass loading coefficient, $\eta_\mathrm{ml}$,
constant, despite they may change with time as the number of massive stars and
the amount of gas left over from star formation decrease.

\section{Results}
\label{sec:results}

In the first part of this section (\S\ref{ssec:proto}), we describe in detail
results for our reference model whose parameters are chosen to represent a
typical SSC. Since the heating efficiency, $\eta_\mathrm{he}$, and the mass
loading factor, $\eta_\mathrm{ml}$, are free parameters, we show results for
three different combinations of them. In section \S\ref{ssec:dep}, we show how
the most important results (the existence of the bimodal regime and the amount
of the accumulated mass) depend on the cluster mass, the cluster radius and the
initial metallicity of stars and the mass-loaded gas $Z_0$.

\subsection{The reference model}
\label{ssec:proto}

We calculate the evolution of a wind driven by a cluster with a stellar mass
$M_\star = 10^6$~M$_\odot$, radius $R_\mathrm{SC} = 3$~pc and initial stellar
metallicity $Z_0 = Z_\odot = 0.02$. We explore three combinations of
$\eta_\mathrm{he}$ and $\eta_\mathrm{ml}$ (see Table~\ref{tab:tabmod}). In
model~A, there is no mass loading and the heating efficiency is 100\%. Model~B
is chosen to be in agreement with
\citet{2007ApJ...669..952S,2009ApJ...700..931S} who have obtained
$\eta_\mathrm{he} \simeq 5\%$, in order to fit the parameters of the compact HII
regions observed around 11 SSCs selected in the central zone of M82. In model C,
the mass loading factor, $\eta_\mathrm{ml} = 19$, is set to give the same value
of $V_{\eta,\infty}$, as in model B, where
\begin{equation}
V_{\eta,\infty} = \left[\frac{2 \eta_\mathrm{he} L_\mathrm{SC}}{(1 +
\eta_\mathrm{ml}) \dot{M}_\mathrm{SC}}\right]^{1/2} .
\label{eq:vinf} 
\end{equation} 
is the adiabatic wind terminal speed corrected for effects of heating efficiency and
mass loading. 

\begin{table}
\begin{tabular}{|c|c|c|c|c|c|c|}
\hline
Model & $\eta_\mathrm{he}$ & $\eta_\mathrm{ml}$ 
& $t_\mathrm{bs}$ [Myr] & $t_\mathrm{be}$ [Myr] &
$M_\mathrm{acc}$ [M$_\odot$] & $M_\mathrm{in}$ [M$_\odot$]\\
\hline
A & $1$    & $0$  & -   & -    & $0$              & $1.8\times 10^5$\\
B & $0.05$ & $0$  & 2.4 & 11.1 & $5.8\times 10^4$ & $1.8\times 10^5$\\
C & $1$    & $19$ & 1.3 & 16.9 & $1.8\times 10^6$ & $3.7\times 10^6$\\
\hline
\end{tabular}
\caption{The reference model calculated with different $\eta_\mathrm{he}$ and
$\eta_\mathrm{ml}$. Columns 4 and 5 denote the beginning and the end of the
period of bimodality ($L_\mathrm{SC} > L_\mathrm{crit}$). Columns 6 and 7 show
the amount of mass accumulated inside the cluster, $M_\mathrm{acc}$, and the
total amount of mass, $M_\mathrm{in}$, supplied into the cluster by stars and
mass loading, respectively.} 
\label{tab:tabmod}
\end{table}

\begin{figure}
\plotone{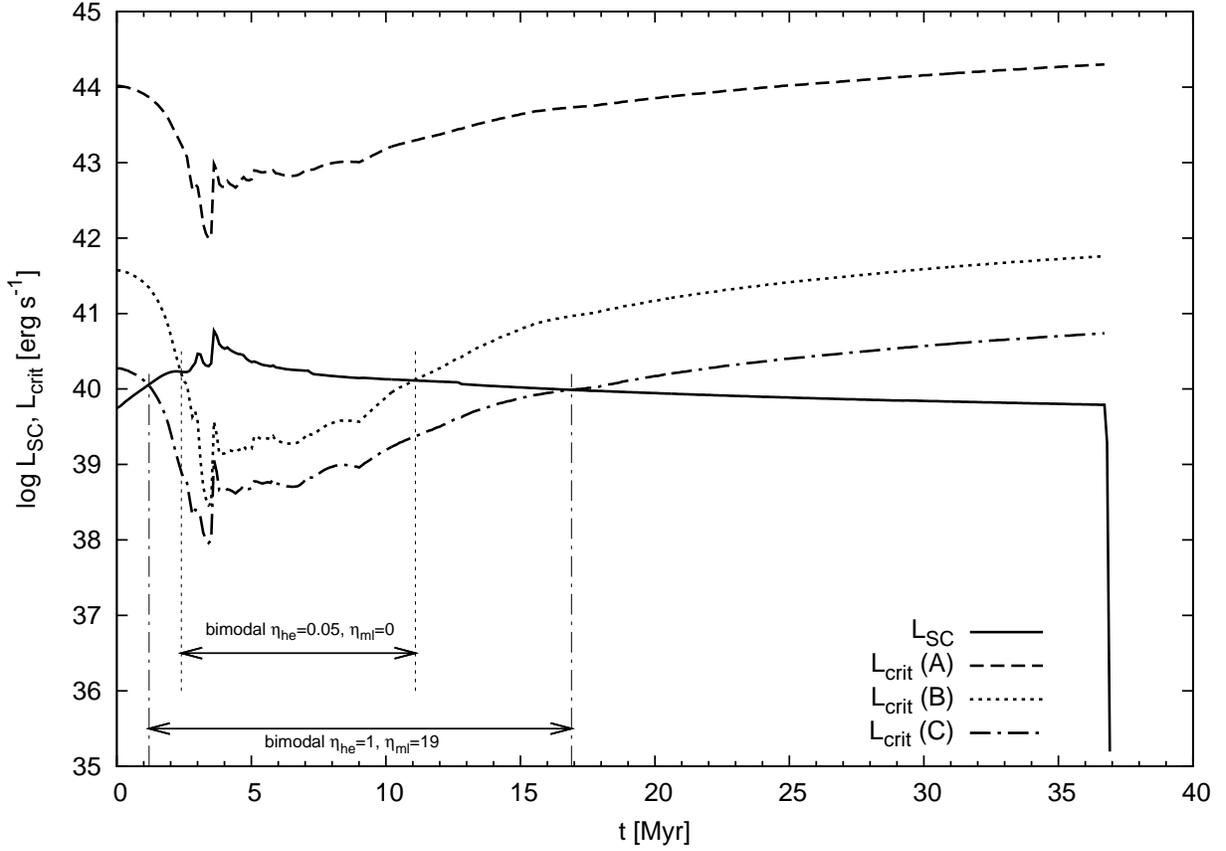}
\caption{The evolution of the critical luminosity, $L_\mathrm{crit}$, for models
A (dashed), B (dotted) and C (dash-dotted).
The $L_\mathrm{crit}$ curves are compared to the cluster mechanical luminosity
$L_\mathrm{SC}$ (solid). Periods during which the cluster evolves in the bimodal
regime are denoted by arrows.}
\label{fig:Lcrit}
\end{figure}

Figure~\ref{fig:Lcrit} compares the time evolution of the critical luminosity,
$L_\mathrm{crit}$, with the star cluster mechanical luminosity, $L_\mathrm{SC}$.
In model~A the star cluster mechanical luminosity is always below the critical
value, $L_\mathrm{SC} < L_\mathrm{crit}$, and thus all gas re-inserted by stars
leaves the cluster as a wind. On the other hand, models~B and C present periods
with $L_\mathrm{SC} > L_\mathrm{crit}$ when clusters evolve in the bimodal
regime. The beginning and the end of these periods are shown in
Table~\ref{tab:tabmod} in columns $t_\mathrm{bs}$ and $t_\mathrm{be}$, respectively. Even
though models~B and C have the same $V_{\eta,\infty}$, the period of
bimodality is longer in model~C. This is because, due to mass loading, the
density of the thermalized plasma is larger in model C and it results in a
higher cooling rate that favors thermal instabilities and mass accumulation.

\begin{figure}
\plotone{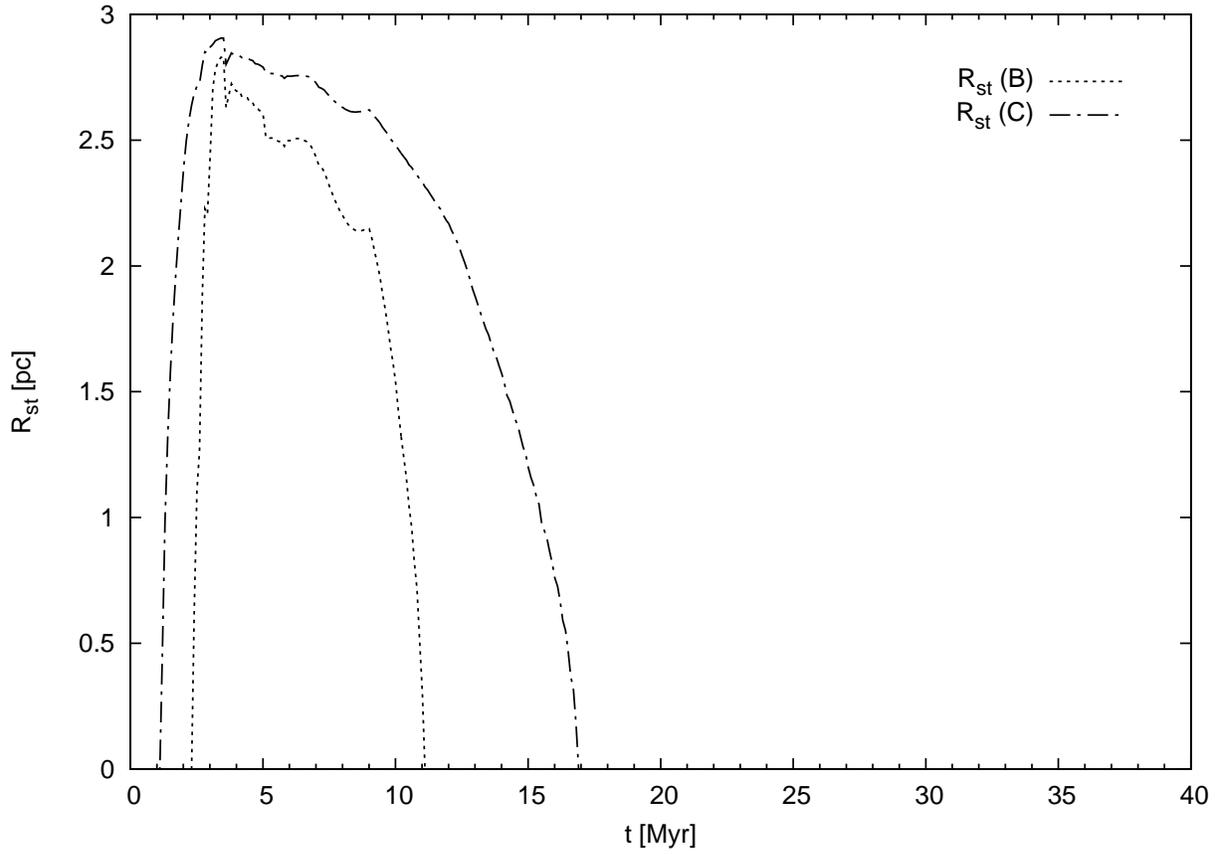}
\caption{The evolution of the stagnation radius, $R_\mathrm{st}$, for models B
(dotted) and C (dash-dotted). The stagnation radius is always zero in model~A.
}
\label{fig:Rst}
\end{figure}

The stagnation radius, $R_\mathrm{st}$, for the three models is shown in
Figure~\ref{fig:Rst}. For model~A, it is always at the cluster center, while in
the other two cases, $R_\mathrm{st}$ reaches a substantial fraction of
$R_\mathrm{SC}$ when the clusters evolve in the bimodal regime. This implies
that the amount of mass accumulated in the central zones of the cluster may be
significant if the heating efficiency is low or the mass loading is large. 
It is because the mass accumulation rate is $\dot{M}_\mathrm{acc} =
\dot{M}_\mathrm{in} (R_\mathrm{st}/R_\mathrm{SC})^3$ where
$\dot{M}_\mathrm{in} = (1 + \eta_\mathrm{ml}) \dot{M}_\mathrm{SC}$ is the rate
at which mass is supplied into the cluster by stars and mass loading.
For example, the amount of the accumulated matter, $M_\mathrm{acc} =
\int_{t_\mathrm{bs}}^{t_\mathrm{be}} \dot{M}_\mathrm{acc} \mathrm{d}t$, is about
one third of the total mass supplied into the cluster, $M_\mathrm{in} =
\int_{t_\mathrm{bs}}^{t_\mathrm{be}} \dot{M}_\mathrm{in} \mathrm{d}t$, in the
case of model~B and about one half of $M_\mathrm{in}$ in the case of model~C
(see Table~\ref{tab:tabmod}).

\begin{figure}
\plotone{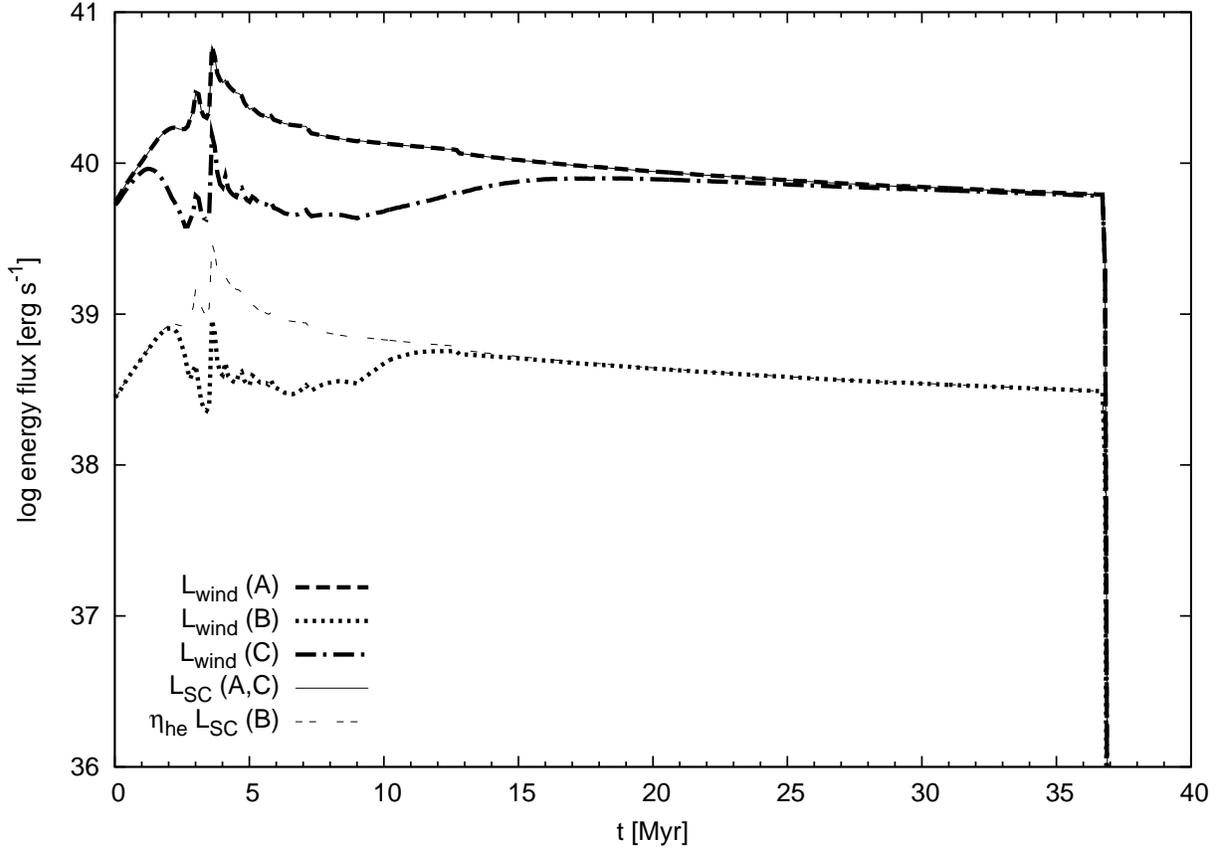}
\caption{The evolution of the wind mechanical output rate, $L_\mathrm{wind}$,
for models A, B and C is shown by thick dashed, dotted and dash-dotted lines,
respectively. The thin solid curve shows the star cluster mechanical luminosity,
$L_\mathrm{SC}$. It overlaps with the thick dashed curve, because for model A,
cooling from the hot wind is negligible and $L_\mathrm{wind} = L_\mathrm{SC}$.
The thin double-dashed line represents the heating efficiency reduced energy
deposition rate, $\eta_\mathrm{he}L_\mathrm{SC}$, for model~B with
$\eta_\mathrm{he}=0.05$.}
\label{fig:Efluxes}
\end{figure}

\begin{figure}
\plotone{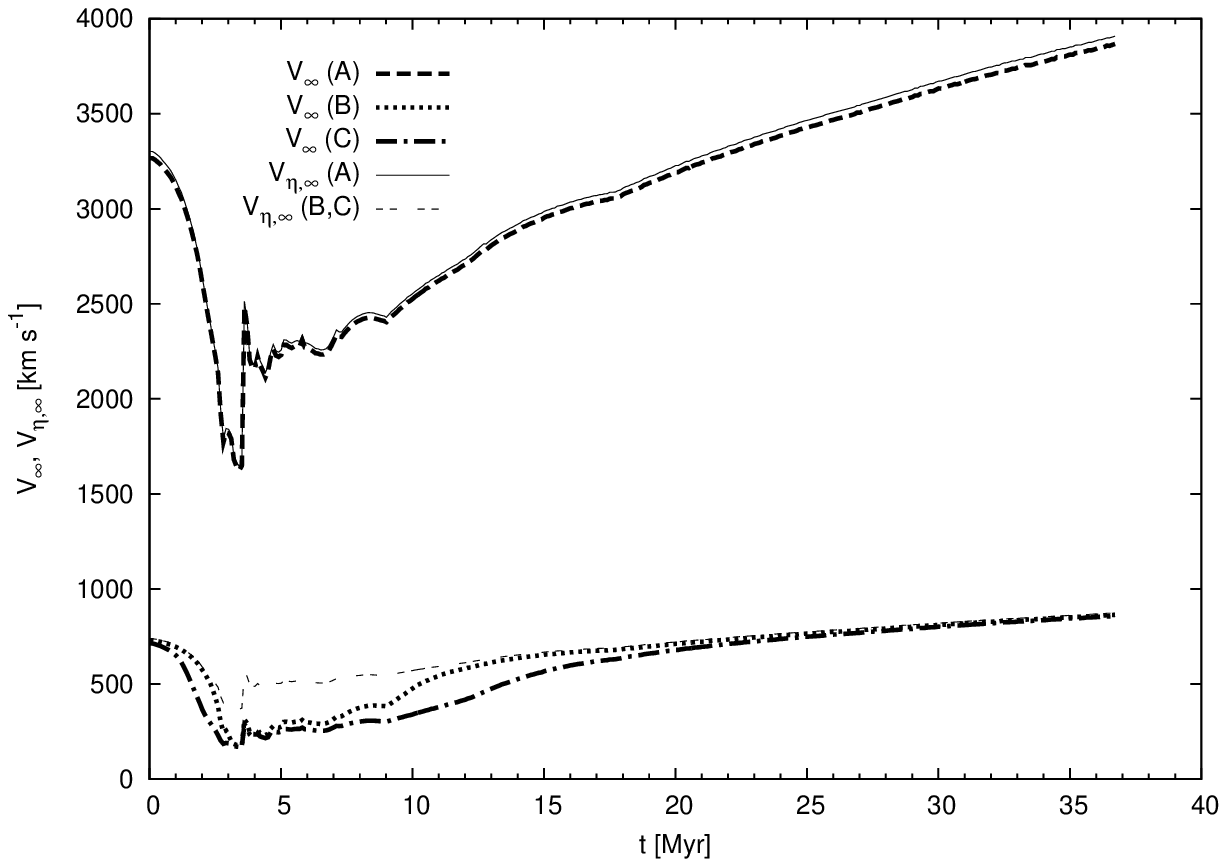}
\caption{
The evolution of wind terminal speed, $V_{\infty}$. Thick dashed, dotted and
dash-dotted lines show the calculated wind terminal speed in the case of models
A, B and C, respectively. Thin lines display $V_{\eta,\infty}$ (see
equation~\ref{eq:vinf}), for model~A (solid) and models B and C (double dashed).
}
\label{fig:vinf}
\end{figure}

Note that strong radiative cooling also affects the star cluster wind mechanical
output rate, $L_\mathrm{wind} = 4 \pi \rho u r^2 (u^2/2 + H)$, where $H$ is the
enthalpy. Figure~\ref{fig:Efluxes} shows that in the bimodal regime it falls
well below the star cluster mechanical luminosity, $L_\mathrm{SC}$ (model C),
and below the heating efficiency reduced star cluster mechanical luminosity,
$\eta_\mathrm{he}L_\mathrm{SC}$ (model B). This implies that the ``true'' energy
output and thus the impact of SSCs on the ambient ISM may be much smaller than
one would expect from star cluster synthetic models like Starburst99. Note
also that the star cluster wind terminal speed is in such cases smaller than
that expected from the star cluster synthetic models, see Figure~\ref{fig:vinf}.
It compares the wind terminal speed $V_\infty$ (measured from semi-analytic
models at $r = 10~R_\mathrm{SC}$) to the heating efficiency and mass loading
corrected adiabatic wind terminal speed $V_\mathrm{\eta,\infty}$. The difference
between the two, significant mainly during bimodality periods, is due to the
radiative energy losses from the wind.

\subsection{Dependence on the stellar cluster parameters}
\label{ssec:dep}

\begin{figure}
\plotone{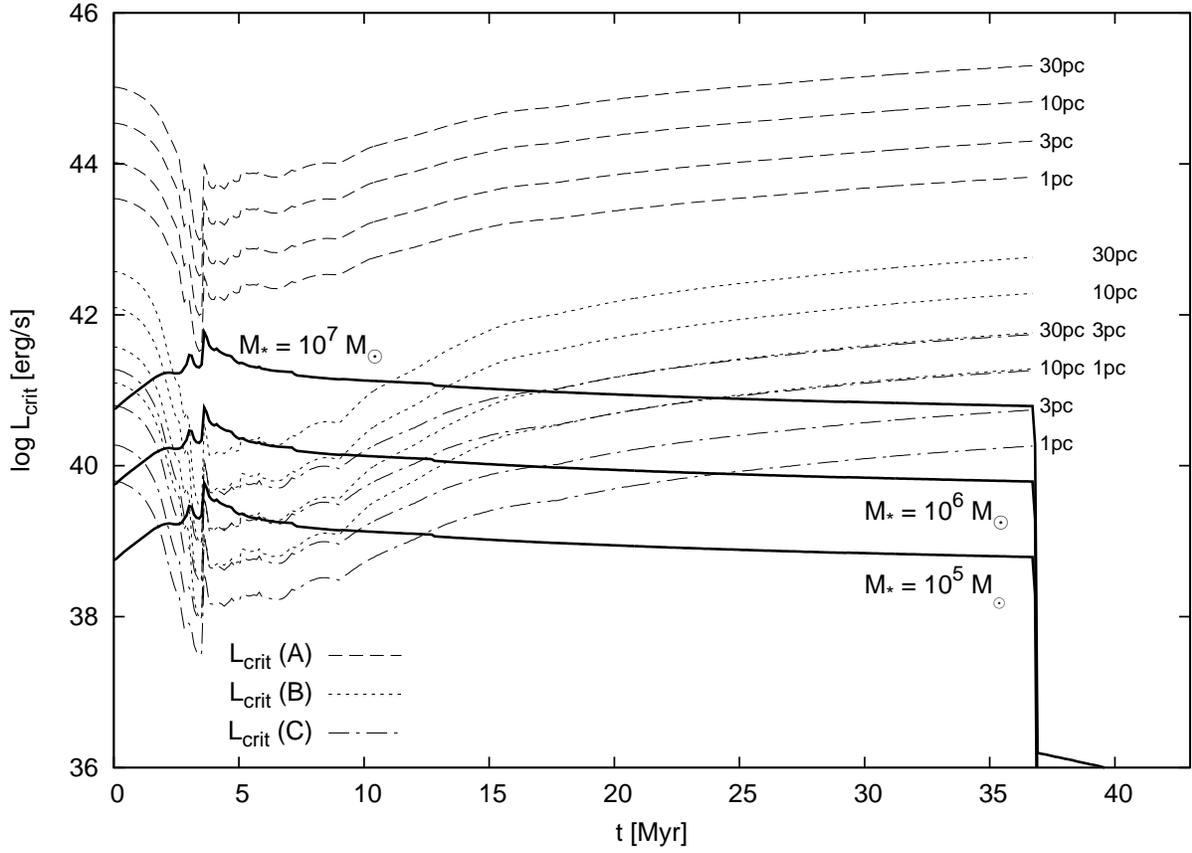}
\caption{Thin lines show the evolution of the critical luminosity,
$L_\mathrm{crit}$, for models A, B and C (the lines have the same meaning as in
Figure \ref{fig:Lcrit}) and different cluster radii, $R_\mathrm{SC}$, marked in
the figure. Thick solid lines display the mechanical luminosity,
$L_\mathrm{SC}$, for clusters with different stellar masses as denoted in the figure.}
\label{fig:MRdep}
\end{figure}

In this section we explore how our results depend on the cluster parameters
running models~A, B and C for clusters with different masses ($M_{\star} = 10^5,
10^6$ and $10^7 \Msol$), different radii ($R_\mathrm{SC} = 1$, $3$,
$10$ and $30$~pc) and taking into consideration the variation of the re-inserted gas
metallicity. The results of the calculations for stellar clusters with different
masses and radii, when the re-inserted and the ablated gas metallicity were
fixed to the solar value are presented in Figure~\ref{fig:MRdep}. This figure
compares the calculated critical mechanical luminosities, $L_\mathrm{crit}$, to
the star cluster mechanical luminosity obtained from the Starburst99 synthetic
model. Figure~\ref{fig:MRdep} shows that clusters with $\eta_\mathrm{he} = 1$
and $\eta_\mathrm{ml} = 0$ never evolve in the bimodal regime. On the other
hand, models with low heating efficiency or large mass loading exhibit periods
of bimodality (see Tables~\ref{tab:MSCdep} and \ref{tab:RSCdep}). In the extreme
cases the amount of mass accumulated inside the cluster, $M_\mathrm{acc}$, may
reach 70\% of the re-inserted and ablated mass, as it is the case when
$\eta_{ml} = 19$ and $M_{\star} = 10^7$\Msol. Note that \citet{2007A&A...471..579W}
derived an approximate analytic formula for $L_\mathrm{crit}$ which predicts
that $L_\mathrm{crit}$ is in direct proportion to the size of the cluster,
$R_\mathrm{SC}$. This is in excellent agreement with our semi-analytic results.
Note also that both $L_\mathrm{SC}$ and $\dot{M}_\mathrm{SC}$ are linearly
proportional to $M_\star$ resulting in $L_\mathrm{crit}$ independent of
$M_\star$. Thus, $L_\mathrm{crit}$ defines the critical cluster mass,
$M_\mathrm{crit}$, and clusters evolve in the bimodal regime if $M_\star >
M_\mathrm{crit}$. This linear dependence may be broken if the cluster IMF varies with
the cluster mass, or if more massive clusters are formed in a different more
abrupt process compared to low mass clusters. However, in this paper we
explore consequences of an abrupt cluster formation with a given IMF.
Discussion of their dependence on the cluster mass exceeds the scope of this
paper.
The results of the calculations for clusters with different masses and radii in
the case when the inserted gas metallicity is solar are summarized in
Tables~\ref{tab:MSCdep} and \ref{tab:RSCdep}. The tables show that even in the
case of low heating efficiency or large mass loading, clusters evolve in the
bimodal regime only for some time, as it was suggested in
\citet{2009ApJ...700..931S}. The length of the period of bimodality and the
amount of accumulated mass are larger for clusters with smaller radii and larger
masses.

\begin{table}
\begin{tabular}{|c|c|c|c|c|c|c|}
\hline
$M_\star$ M$_\odot$ & $\eta_\mathrm{he}$ & $\eta_\mathrm{ml}$ 
& $t_\mathrm{bs}$ [Myr] & $t_\mathrm{be}$ [Myr] &
$M_\mathrm{acc}$ [M$_\odot$] & $M_\mathrm{in}$ [M$_\odot$]\\
\hline
$10^5$ & $1$    & $0$  & -   & -    & $0$ & $1.8\times 10^4$\\
$10^6$ & $1$    & $0$  & -   & -    & $0$ & $1.8\times 10^5$\\
$10^7$ & $1$    & $0$  & -   & -    & $0$ & $1.8\times 10^6$\\
\hline                                 
$10^5$ & $0.05$ & $0$  & 3.1 &  5.1 & $1.1\times 10^3$ & $1.8\times 10^4$\\
$10^6$ & $0.05$ & $0$  & 2.4 & 11.1 & $5.8\times 10^4$ & $1.8\times 10^5$\\
$10^7$ & $0.05$ & $0$  & 1.6 & 17.3 & $1.0\times 10^6$ & $1.8\times 10^6$\\
\hline                                 
$10^5$ & $1$    & $19$ & 2.1 &  9.8 & $8.0\times 10^4$ & $3.7\times 10^5$\\
$10^6$ & $1$    & $19$ & 1.2 & 16.9 & $1.8\times 10^6$ & $3.7\times 10^6$\\
$10^7$ & $1$    & $19$ & 0.0 & 36.8 & $2.8\times 10^7$ & $3.7\times 10^7$\\
\hline
\end{tabular}
\caption{Clusters with different stellar mass, $M_\star$, heating efficiency,
$\eta_\mathrm{he}$, and mass loading $\eta_\mathrm{ml}$. The cluster radius is
$R_\mathrm{SC} = 3$~pc for all these models. Columns 4 -- 7 have the
same meaning as in Table~\ref{tab:tabmod}.}
\label{tab:MSCdep}
\end{table}

\begin{table}
\begin{tabular}{|c|c|c|c|c|c|c|}
\hline
$R_\mathrm{SC}$ [pc] & $\eta_\mathrm{he}$ & $\eta_\mathrm{ml}$ 
& $t_\mathrm{bs}$ [Myr] & $t_\mathrm{be}$ [Myr] 
& $M_\mathrm{acc}$ [M$_\odot$] & $M_\mathrm{in}$ [M$_\odot$] \\
\hline
$1$  & $1$    & $0$  & -   & -    & $0$ & $1.8\times 10^5$ \\
$3$  & $1$    & $0$  & -   & -    & $0$ & $1.8\times 10^5$ \\
$10$ & $1$    & $0$  & -   & -    & $0$ & $1.8\times 10^5$ \\
$30$ & $1$    & $0$  & -   & -    & $0$ & $1.8\times 10^5$ \\
\hline                                                 
$1$  & $0.05$ & $0$  & 2.1 & 13.4 & $8.0\times 10^4$ & $1.8\times 10^5$ \\
$3$  & $0.05$ & $0$  & 2.4 & 11.1 & $5.8\times 10^4$ & $1.8\times 10^5$ \\
$10$ & $0.05$ & $0$  & 2.8 &  9.2 & $3.1\times 10^4$ & $1.8\times 10^5$ \\
$30$ & $0.05$ & $0$  & 3.1 &  5.1 & $1.1\times 10^4$ & $1.8\times 10^5$ \\
\hline
$1$  & $1$    & $19$ & 0.2 & 24.2 & $2.3\times 10^6$ & $3.7\times 10^6$ \\
$3$  & $1$    & $19$ & 1.2 & 16.9 & $1.8\times 10^6$ & $3.7\times 10^6$ \\
$10$ & $1$    & $19$ & 1.8 & 12.4 & $1.3\times 10^6$ & $3.7\times 10^6$ \\
$30$ & $1$    & $19$ & 2.1 &  9.8 & $8.0\times 10^5$ & $3.7\times 10^6$ \\
\hline                                                 
\end{tabular}
\caption{Clusters with different radius, $R_\mathrm{SC}$, heating efficiency,
$\eta_\mathrm{he}$, and mass loading $\eta_\mathrm{ml}$. The cluster
stellar mass is $M_\star = 10^6$~M$_\odot$ for all these models. Columns 4 -- 7
have the same meaning as in Table~\ref{tab:tabmod}.}
\label{tab:RSCdep}
\end{table}

\begin{figure}
\plotone{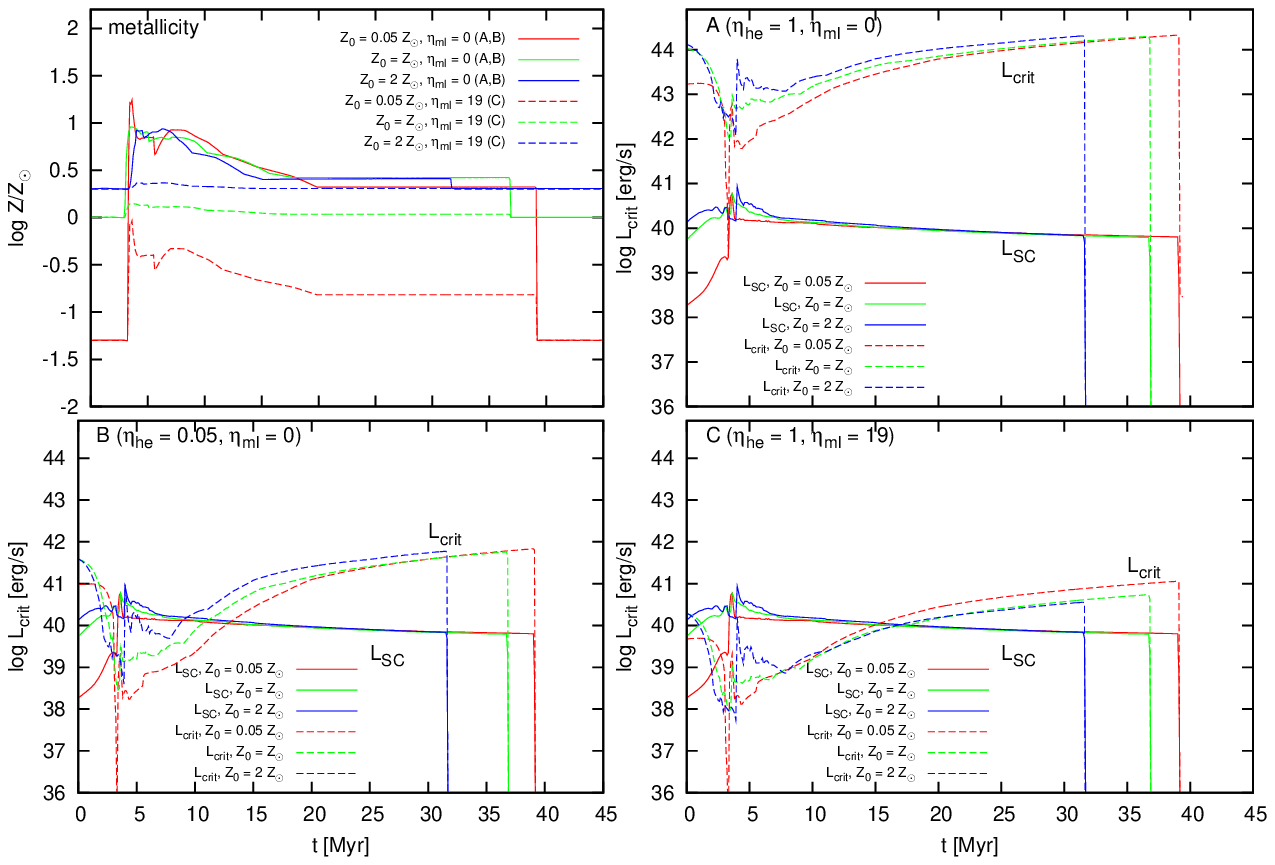}
\caption{Dependence on the stellar metallicity and the metallicity of the mass
loaded gas $Z_0$. Top left panel shows the metallicity of the supplied gas, $Z$,
for models with $\eta_\mathrm{ml} = 0$ (solid, models A and B) and
$\eta_\mathrm{ml} = 19$ (dashed, model~C). Other panels compare the evolution of
$L_\mathrm{SC}$ and $L_\mathrm{crit}$ with different $Z_0$ for models~A (top
right), models~B (bottom left) and models~C (bottom right). In all panels, colors of
curves represent $Z_0$: $Z_0 = 0.05$~Z$_\odot$ (red), $Z_0 =$~Z$_\odot$ (green) and $Z_0 =
2.0$~Z$_\odot$ (blue).}
\label{fig:Lcrit_zdep}
\end{figure}

Another parameter which may affect properties of the star cluster driven
outflows is the re-inserted gas metallicity. In the case of instantaneous star
formation, the metallicity of the re-inserted matter changes a lot, as it is
shown in Figure~\ref{fig:chem}. This should change the cooling rate and thus the
critical mechanical luminosity, $L_\mathrm{crit}$, significantly
\citep{2005ApJ...628L..13T}. In order to explore how our results depend on this
parameter, we have varied the initial stellar and the loaded gas metallicity,
$Z_0$, in our reference models A, B and C. Three different values of $Z_0$ were
used for the calculations: $Z_0 = 0.05 Z_{\odot}$, $Z_0 = Z_{\odot}$ and $Z_0 =
2.0 Z_{\odot}$. The top left panel in Figure~\ref{fig:Lcrit_zdep} shows the
trends of the wind metallicity, $Z$, calculated from equation \ref{eq:zmix}. In
all cases without mass loading (solid lines in Figure~\ref{fig:Lcrit_zdep}) the
metallicity of the thermalized plasma grows rapidly after the first supernova
explodes, reaches about 10 times the solar value, and then decreases gradually
reaching about 3 times the solar value after $\sim 20$~Myr. In the case with mass
loading, the maximum metallicity never reaches 10 times the solar value. This is
because in this case the re-inserted matter mixes continuously with a large
amount of the ablated gas. The calculated critical luminosities,
$L_\mathrm{crit}$, are then compared with the star cluster mechanical
luminosities, $L_\mathrm{SC}$ (top right, bottom left and bottom right panels in
Figure~\ref{fig:Lcrit_zdep} for cases A, B and C, respectively). Models without
mass loading and $\eta_\mathrm{he} = 1$ never enter the bimodal regime (see top
right panel). Note that relative abundances of species in the re-inserted matter
differs from solar values. This implies that the cooling function using scaled
solar composition $Z$ may give somewhat different cooling rates that that
calculated from individual species separately. This, however, does not change
our results significantly, since the main coolants (C and O) are also dominant
ingredients of $Z$.

\begin{table}
\begin{tabular}{|c|c|c|c|c|c|c|}
\hline
$Z_0$ [Z$_\odot$] & $\eta_\mathrm{he}$ & $\eta_\mathrm{ml}$ 
& $t_\mathrm{bs}$ [Myr] & $t_\mathrm{be}$ [Myr] 
& $M_\mathrm{acc}$ [M$_\odot$] & $M_\mathrm{in}$ [M$_\odot$] \\
\hline
$0.05$  & $1$    & $0$  & -   & -    & $0$ & $1.8\times 10^5$ \\
$1.0$   & $1$    & $0$  & -   & -    & $0$ & $1.8\times 10^5$ \\
$2.0$   & $1$    & $0$  & -   & -    & $0$ & $1.8\times 10^5$ \\
\hline                                 
$0.05$  & $0.05$ & $0$  & 3.1 & 12.9 & $7.5\times 10^4$ & $1.8\times 10^5$ \\
$1.0$   & $0.05$ & $0$  & 2.4 & 11.1 & $5.8\times 10^4$ & $1.8\times 10^5$ \\
$2.0$   & $0.05$ & $0$  & 1.9 &  9.3 & $3.7\times 10^4$ & $1.8\times 10^5$ \\
\hline
$0.05$  & $1$    & $19$ & 2.8 & 15.0 & $1.6\times 10^6$ & $3.6\times 10^6$ \\
$1.0$   & $1$    & $19$ & 1.2 & 16.9 & $1.8\times 10^6$ & $3.7\times 10^6$ \\
$2.0$   & $1$    & $19$ & 0.4 & 17.0 & $1.8\times 10^6$ & $3.6\times 10^6$ \\
\hline                                 
\end{tabular}
\caption{Models with different stellar metallicity, $Z_0$, heating efficiency,
$\eta_\mathrm{he}$, and mass loading $\eta_\mathrm{ml}$. Other cluster
parameters are $R_\mathrm{SC} = 3$~pc and $M_\star = 10^6$~M$_\odot$. Columns 4
-- 7 have the same meaning as in Table~\ref{tab:tabmod}.}
\label{tab:zdep}
\end{table}

\section {Conclusions}
\label{sec:conclusions}

We used our semi-analytic spherically-symmetric code together with the stellar
population synthesis model Starburst99 to study the time evolution of Super
Star Cluster winds. 

Two physical processes which could affect the hydrodynamics of the star cluster
winds significantly and cannot be studied in the semi-analytic approach
in details, the heating efficiency and mass loading, are parameterized with two
constant parameters $\eta_\mathrm{he}$ and $\eta_\mathrm{ml}$. We also search
how our major results depend on the metallicity of the re-inserted matter.

The calculations show that strong radiative cooling becomes a crucial issue when
the wind is mass loaded or the thermalization efficiency (and thus the fraction
of the star cluster mechanical luminosity which drives the outflow) is small. In
these cases (our reference models C and B, respectively) the evolutionary tracks
of the star cluster winds show periods of bimodality. During these periods only
some fraction of the re-inserted and loaded gas leaves the cluster as a wind.
The rest of the re-inserted matter cools down rapidly, becomes thermally
unstable and is accumulated in the central region of the cluster. The duration
of these periods depends on the star cluster parameters $\eta_\mathrm{he}$ and
$\eta_\mathrm{ml}$. Periods of bimodality are longer in the case of more massive
clusters with smaller radii. However, they become progressively shorter as the
mass loading drops or the heating efficiency grows. The bimodal regime vanishes
in the cases when heating efficiency is large and mass loading is insignificant.
In the simulations which include mass loading, the stellar metallicity does not
affect significantly neither the duration of the bimodal regime nor the amount
of re-inserted mass which accumulates inside the cluster. Models with low
heating efficiency are more sensitive to the metallicity of the re-inserted
matter.

We conclude that the second stellar generation may be formed in massive and
compact stellar clusters from thermally unstable parts of stellar winds and the
mass loaded gas in their central parts. Low heating efficiency $\eta_\mathrm{he}
= 0.05$ leads to the second stellar generation heavily enriched with He-burning
products. However, its total mass is a few percent of the first generation only.
High value of mass loading $\eta_\mathrm{ml} = 19$ results in the massive
second stellar generation, however, its metallicity is only slightly higher
than that of the first generation.

\acknowledgments 
We thank our anonymous referee for valuable comments and suggestions.
We thank James E.~Dale for careful reading of the text and his useful
suggestions. This study has been supported by CONACYT - M\'exico, research
grants 60333 and 131913 and the Spanish Ministry of Science and Innovation under
the collaboration ESTALLIDOS (grant AYA2010-21887-C04-04) and Consolider-Ingenio
2010 Program grant CSD2006-00070: First Science with the GTC. RW and JP
acknowledge support from the Institutional Research Plan AV0Z10030501 of the
Academy of Sciences of the Czech Republic and project LC06014 Centre for
Theoretical Astrophysics of the Ministry of Education, Youth and Sports of the
Czech Republic.

\bibliographystyle{aa}
\bibliography{clevol}

\end{document}